\documentclass[twocolumn,aps,prl,showpacs,groupedaddress,superscriptaddress]{revtex4-2}
\usepackage[latin9]{inputenc}
\usepackage{color}
\usepackage{bm}
\usepackage{amsmath}
\usepackage{amssymb}
\usepackage{graphicx}
\usepackage[bookmarks=true,colorlinks,linkcolor=blue,urlcolor=blue,citecolor=blue]{hyperref}

\makeatletter

\usepackage{bm}
\usepackage{lipsum}
\usepackage{mathbbol}

\makeatother

\begin{document}
\title{Optical phonons coupled to a Kitaev spin liquid}

\author{A. Metavitsiadis}
\email{a.metavitsiadis@tu-bs.de}
\affiliation{Institute for Theoretical Physics, Technical University
Braunschweig, D-38106 Braunschweig, Germany}

\author{W. Natori}
\email{w.natori@imperial.ac.uk}
\affiliation{Blackett Laboratory, Imperial College London, London SW7 2AZ,
United Kingdom}
\affiliation{Institut Laue-Langevin, BP 156, 41 Avenue des Martyrs, 38042 Grenoble Cedex 9, France}

\author{J. Knolle}
\email{j.knolle@tum.de}
\affiliation{Department of Physics TQM,
Technical University of Munich, 85748 Garching, Germany}
\affiliation{Munich Center for Quantum Science and Technology (MCQST),
Schellingstrasse 4, D-80799 M\"unchen, Germany}
\affiliation{Blackett Laboratory, Imperial College London, London SW7 2AZ,
United Kingdom}

\author{W. Brenig}
\email{w.brenig@tu-bs.de}
\affiliation{Institute for Theoretical Physics, Technical University
Braunschweig, D-38106 Braunschweig, Germany}

\date{\today}
\begin{abstract}
Emergent excitation continua in frustrated magnets are a fingerprint of
fractionalization, characteristic of quantum spin-liquid states.  Recent
evidence from Raman scattering for a coupling between such continua and
lattice degrees of freedom in putative Kitaev magnets
\cite{PhysRevLett.114.147201, PhysRevB.95.174429, PhysRevB.100.134419,
PhysRevB.101.045419,Wulferding2020,wang2020range} may provide insight into the nature of
the fractionalized quasiparticles. Here we study the renormalization of
optical phonons coupled to the underlying $\mathbb{Z}_{2}$ quantum
spin-liquid. We show that phonon line-shapes acquire an asymmetry, observable
in light scattering, and originating from two distinct sources, namely the
dispersion of the Majorana continuum and the Fano effect. Moreover, we find
that the phonon life-times increase with increasing temperature due to
thermal blocking of available phase space. Finally, in contrast to low-energy
probes, optical phonon renormalization is rather insensitive to thermally
excited gauge fluxes and barely susceptible to external magnetic fields.
\end{abstract}
\maketitle

\emph{Introduction.--}  There is an ongoing pursuit of
the signatures of the elusive quantum spin-liquid (QSL) state of matter
\cite{0034-4885-80-1-016502, RevModPhys.89.025003,
doi:10.1146/annurev-conmatphys-031218-013401}.  The difficulty to identify
such states is due to the fact that they do not break any symmetries and
lack conventional local order parameters of magnetic or related nature down
to zero temperature. Recently, QSLs with a $\mathbb{Z}_{2}$ gauge structure
may have actually come close to material realization, motivated by the exact
solution of the famous Kitaev spin model (KSM) with compass-exchange on the
two-dimensional (2D) honeycomb lattice \cite{Kitaev20062}. In this model,
spins fractionalize into static $\mathbb{Z}_{2}$ gauge fluxes and itinerant
Majorana fermions, with a gapless QSL ground state. In external magnetic
fields the KSM opens a gap and displays chiral Majorana edge-modes. Variants
and generalizations of the KSM in 1D \cite{PhysRevB.93.214425,
PhysRevLett.98.087204, Wu20123530, PhysRevB.96.041115, PhysRevB.99.205129,
PhysRevB.99.224418, metavitsiadis2020flux}, 2D \cite{Yang2007}, and 3D
\cite{PhysRevLett.113.197205, OBrien2016, PhysRevB.96.125124}, as well as for
spins larger than $1/2$ \cite{PhysRevB.78.115116, Rousochatzakis2018,
PhysRevLett.123.037203} have been considered.

Mott-insulators with strong spin-orbit coupling (SOC) are promising materials
to realize the KSM \cite{Khaliullin2005, PhysRevLett.102.017205,
PhysRevLett.105.027204, RevModPhys.87.1}. However, residual non-Kitaev
exchange interactions remain an issue, with all current systems under
consideration eventually displaying magnetic order at low temperatures. From
a present perspective \cite{Trebst2017, Winter2017, 033117-053934,
doi:10.7566/JPSJ.89.012002, doi:10.1146/annurev-conmatphys-031218-013401},
$\alpha$-RuCl$_3$, either above its ordering temperature, or with magnetic
order suppressed by external magnetic fields, is one of the prime candidates
under scrutiny for $\mathbb{Z}_{2}$ QSL physics. Recent thermal Hall effect
measurements in $\alpha$-RuCl$_3$, suggest half-integer quantization plateaus
\cite{Kasahara2018} which are consistent with Majorana edge states, including
a field-angular variation of the topological Chern number identical to that
of the Kitaev QSL \cite{yokoi2020halfinteger}, and a bulk-boundary
correspondence claimed in specific heat measurements
\cite{tanaka2020thermodynamic}.
In addition to edge transport, a multitude of bulk spectroscopic probes have
been invoked, aiming to identify continua characteristic of the fractional
Majorana excitations. This pertains to inelastic neutron scattering
\cite{Banerjee2016, Banerjee2016a, Banerjee2018, knolle2014dynamics, smith2015neutron, knolle2018dynamics} and local resonance
techniques \cite{Baek2017, Zheng2017}, as well as to magnetic Raman scattering
\cite{PhysRevLett.113.187201, Nasu2016,perreault2015theory}.

An interesting open question is whether the coupling of Majorana fermions of the putative KSM to other degrees of freedom can be used to provide signatures of their existence. Coupling to
phonons \cite{PhysRevX.8.031032, PhysRevLett.121.147201, PhysRevB.101.035103,
PhysRevResearch.2.033180} can induce characteristic renormalizations,
examples of which seem to have been observed recently for acoustic phonons
\cite{Li2020}. Regarding optical phonons, Raman scattering
\cite{PhysRevLett.114.147201, PhysRevB.95.174429, PhysRevB.100.134419,
PhysRevB.101.045419, Wulferding2020,wang2020range} has provided early on evidence for Raman active
phonons with Fano line-shapes, overlapping with the magnetic Raman continuum
\cite{PhysRevLett.113.187201, Nasu2016}. This has been speculated to be a
signature of renormalization of optical phonons by Majorana fermions but up to now a microscopic description was missing.

Here, we provide a theory of optical phonons and coupled to a
KSM. We describe the microscopic details of the coupling and evaluate the
phonon self-energy. The phonon renormalization versus energy and temperature
is studied, the relative importance of Majorana and flux excitations is
discussed, and the implications for Raman scattering are clarified. We find
convincing qualitative agreement  with experimental data, which points to an intriguing interpretation of a Majorana scattering induced optical phonon renormalization in the candidate material $\alpha$-RuCl$_3$.

\emph{Optical-phonon Majorana mixing.--} Here, we consider phonons
of a 2D Honeycomb lattice coupled magnetoelastically to the
Kitaev QSL. The total Hamiltonian of the system reads $H=H_{P}+H_{K}+H_{KP}$
where $H_{P}$ stands for the quadratic free phonon contribution 
$H_{P}=\sum_{\mathbf{q}m}\left(\omega_{m\mathbf{q}}+\frac{1}{2}\right)b_{m\mathbf{
q}}^{\dagger}b_{m\mathbf{q}}$
with $b_{m\mathbf{q}}^{\dagger}$, $b_{m\mathbf{q}}$ being bosonic
creation annihilation operators, respectively, at momentum $\mathbf{q}$
for the mode $m$, and $\omega_{m\mathbf{q}}$ the corresponding energy.

\begin{figure}[t!]
\centering{}\includegraphics[width=0.9\columnwidth]{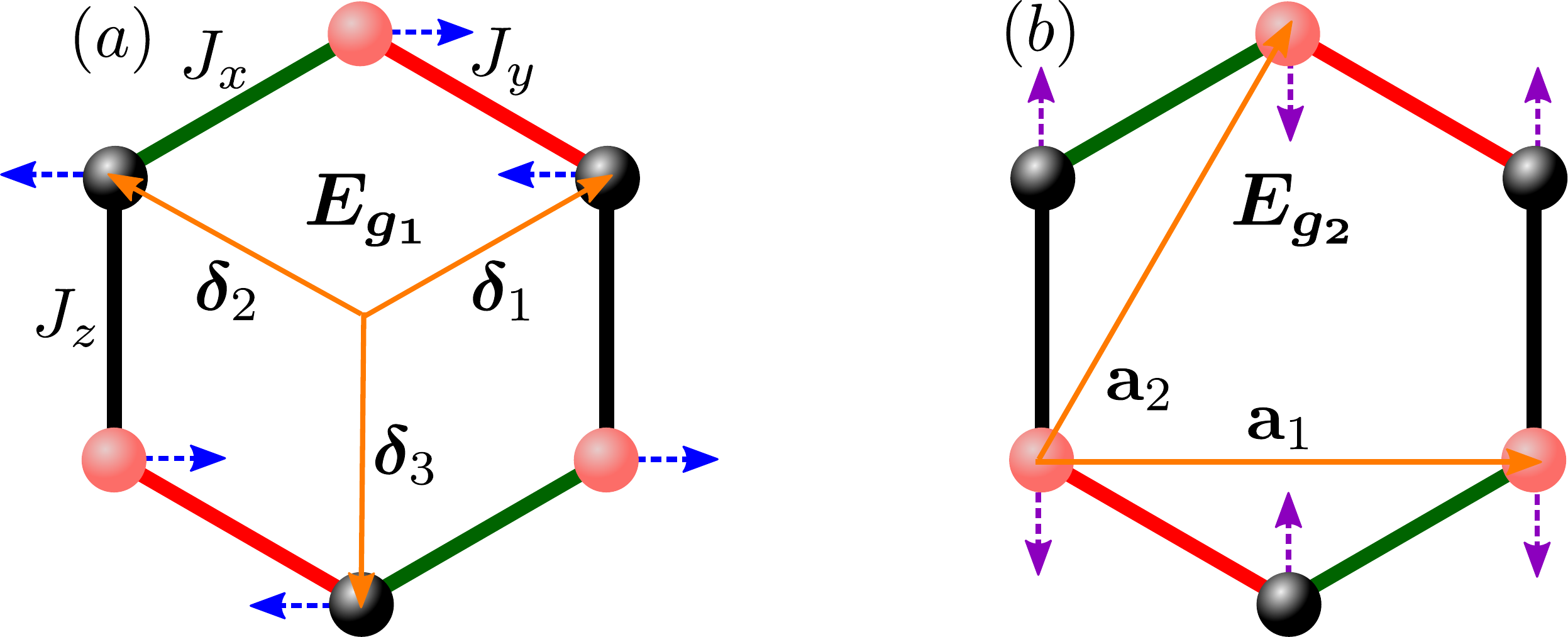}
\caption{Kitaev Hexagons with compass magnetic interactions $J_{x,y,z}$ along
the directions $\bm{\delta}_{1}=\frac{a}{2}(+\sqrt{3}\bm{e}_{x}+\bm{e}_{y})$,
$\bm{\delta}_{2}=\frac{a}{2}(-\sqrt{3}\bm{e}_{x}+\bm{e}_{y})$, and
$\bm{\delta}_{3}=-a\bm{e}_{y}$, respectively. In addition the lattice
vectors of the triangular Bravais lattice are shown $\mathbf{a}_{1}=a
\sqrt{3}\bm{e}_{x}$, $\mathbf{a}_{2}=\frac{a\sqrt{3}}{2}(\bm{e}_{x}+
\sqrt{3}\bm{e}_{y})$. 
The distortion of the lattice due to the two optical modes $E_{g_{1}} (a)$ and 
$E_{g_{2}} (b)$ is indicated by the small blue and purple arrows, respectively, 
with a dashed shaft. \label{fig:model}} 
\end{figure}

The magnetic degrees of freedom are described by the Kitaev spin Hamiltonian
$H_{K}=\sum_{\mathbf{r},\bm{\delta}}J_{\bm{\delta}}S_{\mathbf{r}}^{d_{\bm{\delta}}}
S_{\mathbf{r}-\bm{\delta}}^{d_{\bm{\delta}}}
$, \cite{Kitaev20062} (see also Fig.~\ref{fig:model}). 
$\mathbf{r}=l_{1}\mathbf{a}_{1}+l_{2}\mathbf{a}_{2}$, with $l_{1,2} = 0,
1,\ldots,L-1$, refers to a triangular Bravais lattice of linear dimension $L$,
and $\bm{\delta}$ sets the location of the basis of the honeycomb lattice,
with $N=2L^{2}$ sites. The components $d_{\bm{\delta}}$ of the spin-1/2
operators $S$ assume the values $d_{\bm{\delta}}=x,y,z$ depending on the
$\bm{\delta}$-vector, while $J_{\bm{\delta}}$ are the Kitaev interactions. We
consider the isotropic case $J_{\bm{\delta}}=J$ and set $\hbar,k_{B}=1$.

Following the literature \cite{Kitaev20062,PhysRevLett.98.087204}, we map the
spin model onto one for two species of Majorana fermions $c$ and $\bar{c}$,
with $\{c_{i}, c_{j}\} =2 \delta_{ij}= \{\bar{c}_{i}, \bar{c}_{j}\}$ and
$\{c_{i},\bar{c}_{j}\}=0$. This mapping renders Majorana fermions of, e.g.,
$c$-type itinerant, while the other type pairs into a static $\mathbb{Z}_2$
gauge fields $\eta=\pm 1$ along, e.g., the $\bm{\delta}_3$ direction. The
gauge field generates a conserved flux, equivalent to a macroscopic number
of conserved local operators in the spin language \cite{Kitaev20062,
PhysRevLett.98.247201}. In the Majorana representation, $H_K$ reads
\begin{equation}
H_{K}= J \sum_{\mathbf{r},\bm{\delta}}  h_{\bm{\delta}}(\mathbf{r})~,~~
h_{\bm{\delta}}(\mathbf{r})=-\frac{i}{4}\eta_{\bm{\delta}}(\mathbf{r})
c_{\mathbf{r}}c_{\mathbf{r}-\bm{\delta}}'~,
\label{eq:Hd}
\end{equation}
where the gauge field $\eta_{\bm{\delta}}$ acquires the values
$\eta_{\bm{\delta}_{1}}(\mathbf{r})=\eta_{\bm{\delta}_{2}}(\mathbf{r})=1$,
and $\eta_{\bm{\delta}_{3}}(\mathbf{r})=\pm1$. Primed Majorana fermions
reside on the basis sites. The ground state resides within the uniform gauge
sector, which is separated from other sectors by a gap $\Delta \approx
0.065J$. At finite temperature $T$ fluxes become thermally excited and proliferate in a narrow range near a very low $T^{*} \approx
0.012J$. For several observables the emergent disorder introduced by the
visons has been shown to be of physical significance
\cite{PhysRevB.96.041115, PhysRevB.96.205121, PhysRevB.99.075141,
PhysRevB.101.035103}. For the present case of interest, i.e., optical phonons,
we show that gauge excitations imply only negligible quantitative
modifications.

We focus on magnetoelastic coupling between spins and lattice degrees of
freedom, i.e., on the leading order variation $J(\mathbf{u}_{\mathbf{r}} -
\mathbf{u}_{\mathbf{r} - \bm{\delta}}^{\prime}) \approx J+\nabla
J\cdot(\mathbf{u}_{\mathbf{r}}- \mathbf{u}_{\mathbf{r} -
\bm{\delta}}^{\prime})$ of the exchange with respect to lattice deformations
$\mathbf{u}_{\mathbf{r}}$ at site ${\bf r}$.  The lattice distortions in
Fourier space, $\mathbf{u}_{\mathbf{q}}^{(\prime)}=
\sum_{\mathbf{r}}e^{-i\mathbf{q}\cdot[\mathbf{r}(+\bm{\delta}_1)]}
\mathbf{u}_{\mathbf{r}}^{(\prime)} / \sqrt{N}$, are quantized in terms of
phonon normal modes $\mathbf{u}_{\mathbf{q}}^{(\prime)} = \sum_m
(\gamma_{m\mathbf{q}}^{x(\prime)} \hat{e}_{x}
+\gamma_{m\mathbf{q}}^{y(\prime)} \hat{e}_{y}) / \sqrt{2M
\omega_{m\mathbf{q}}} \hphantom{a} \mathcal{B}_{m \mathbf{q}}$, with
$\mathcal{B}_{m\mathbf{q}}=b_{m\mathbf{q}}+b_{m,-\mathbf{q}}^{\dagger}$
comprising the annihilation and creation operators of mode $m$ at momentum
$\pm\mathbf{q}$ and energy $\omega_{m\mathbf{q}}$ with coefficients of the
polarization vectors $\gamma$ and $\gamma^{\prime}$ and $M$ is of the order
of the ruthenium mass. Using this, the Majorana phonon coupling reads
\begin{equation}
H_{KP}=\sum_{m\mathbf{q}}\mathcal{B}_{m\mathbf{q}}\mathcal{H}_{m,-\mathbf{q}}~,
~~\mathcal{H}_{m\mathbf{q}}=\sum_{\bm{\delta}}
\Lambda_{m\mathbf{q}}^{\bm{\delta}}h_{\bm{\delta};\mathbf{q}},
\label{eq:HKP}
\end{equation}
where the form factor of the coupling is encoded in $\Lambda_{m \mathbf{q}}^{
\bm{\delta}}$.

While Eq. (\ref{eq:HKP}) applies to acoustic, as well as to optical phonons,
we focus on the $E_{g_{1}}$ and $E_{g_{2}}$ optical modes, observed in Raman
experiments. These are of particular interest, since allegedly, they overlap
with the Majorana continuum. For Raman scattering it is safe to consider
$q\rightarrow 0$ only and we drop all $\mathbf{q}$-labels
hereafter. In the supplemental material we treat also $q\neq 0$ (and
moreover weak magnetic fields) \cite{supplement}.  The phonon energies are
$\omega_1 \equiv \omega_{g_{1}}\approx116cm^{-1}\approx1.9J$ and $\omega_2
\equiv \omega_{g_{2}}\approx165cm^{-1}\approx2.6J$, where we assume a Kitaev
coupling of $J\approx90K$. The vibrational pattern of the two modes
\cite{PhysRevB.95.174429} is shown in Fig.~\ref{fig:model}. In terms of
Eq. (\ref{eq:HKP}) the lattice modulations imply $\bm{\Lambda}_{m}
=\Lambda_{m}[1,-1^{m},\lambda_{m}]$, with $m=1,2$, and $\lambda_{m}$ denotes
a possible anisotropy between the $\bm{\delta}_{1,2}$ and $\bm{\delta}_{3}$
directions.  The magnitude of $\Lambda_{m}$ can be assumed to be in the
perturbative regime \cite{PhysRevB.101.035103}. Note, a detailed 
microscopic derivation of the spin-phonon coupling including the effect of spin-orbit effects beyond the pure Kitaev model is given in the supplementary material \cite{supplement}, see also Refs.~\cite{Rau2014, 
Natori2019A, Kugel1982}. 

The coupling of the phonons to the fractionalized magnetic 
excitations, Eq.~\eqref{eq:HKP}, leads to a renormalization of the 
bare phonon propagators, 
$D_{m}^0(z) = 2\omega_{m}/(z^{2}-\omega_{m}^{2})$, 
with $z=\omega + i0^+$, the frequency dependence.  
The dressed phonon propagators can then be determined by the $2\times2$ 
self-energy matrix $\bm{\Sigma}(z)$ and the 
corresponding Dyson's equation,  
\begin{equation}
\mathbf{D}(z)\approx [\mathbf{D}_{0}^{-1}(z) -\bm{\Sigma}(z)]^{-1},~
\bm{\Sigma}_{mm'} = 
\langle\langle\mathcal{H}_{m};\mathcal{H}_{m'}^{\dagger}\rangle\rangle , 
\label{eq:SE}
\end{equation}
with $[\mathbf{D}_0(z)]_{mm'} = \delta_{mm'}D_{m}^0(z)$ 
while the  double brackets denote the Green's function.   
A central goal of the paper is to evaluate the self-energy 
in Eq.~\eqref{eq:SE}. We
do this in two ways: first analytically, assuming a uniform gauge 
field configuration in Eq.~\eqref{eq:Hd},  and second numerically 
by considering a numerical random averaging over disordered 
configurations of the gauge field $\eta$. While the former 
approach is justified at temperatures ranging from 0 
up to $T^{*}$, the latter is valid at temperatures higher 
than the flux gap $T\gtrsim\Delta$ 
\cite{PhysRevB.96.205121,PhysRevB.99.075141,PhysRevB.101.035103}. 

\emph{Phonon self-energy: uniform gauge.--} At low temperatures, $T\lesssim
T^{*}$, it can be assumed, that the system acquires a uniform gauge
configuration, $\eta_{\bm{\delta}}=1$, allowing for the analytical
calculation of the phonon self-energy. First, the Kitaev terms in
Eq.~\eqref{eq:Hd} can be brought to a diagonal form by going to reciprocal
space $\mathbf{k} = k_1\mathbf{G}_{1}+k_{2}\mathbf{G}_{2}$, where
$\mathbf{G}_{1}=\frac{1}{3a}(\sqrt{3}\bm{e}_{x}-\bm{e}_{y})$ and
$\mathbf{G}_{2}=\frac{2}{3a}\bm{e}_{y}$ are the reciprocal lattice vectors,
i.e., $\mathbf{a}_i \cdot \mathbf{G}_j=\delta_{i,j}$ for $i(j)=1,2$.  The
coefficients $k_{1,2}$ are set to antiperiodic boundary conditions
$k_{j}=2\pi(l_{j}+\frac{1}{2})/L$, to allow all Majoranas to pair
into complex fermions.  The
Fourier transform of the Majoranas reads
$c_{\mathbf{k}}= \sum_{\mathbf{r}}e^{-i\mathbf{k}
\cdot\mathbf{r}}c_{\mathbf{r}}/\sqrt{2N}$ such that
$\{c_{\mathbf{k}},c_{\mathbf{k}'}^{\dagger}\}=
\delta_{\mathbf{k},\mathbf{k}'}$, and similarly for the $c'$ operators.

In the diagonal complex fermion basis, $\Psi_{\mathbf{k}}^{\dagger} =
(d_{1,\mathbf{k}}^{\dagger}, d_{2,\mathbf{k}}^{\dagger})$, $H_K$ and
$\mathcal{H}_{m}$ of Eq. \eqref{eq:Hd} and \eqref{eq:HKP} read $H_K =
\frac{1}{2}\sum_{\mathbf{k}} \Psi_{\mathbf{k}}^{\dagger} E_{\mathbf{k}}
\Psi_{\mathbf{k}}$ and $\mathcal{H}_m = \frac{1}{2}\sum_{\mathbf{k}}
\Psi_{\mathbf{k}}^{\dagger} V_{m;\mathbf{k}} \Psi_{\mathbf{k}}$ with
\begin{equation} 
 E_{\mathbf{k}} = 
\left(\begin{array}{cc}
\epsilon_{\mathbf{k}} & 0\\
0 & -\epsilon_{\mathbf{k}}
\end{array}\right), 
\hphantom{aa}
V_{m;\mathbf{k}} = 
\left(\begin{array}{cc}
g_{m;\mathbf{k}}^{\prime} & ig_{m;\mathbf{k}}^{\prime\prime}\\
-ig_{m;\mathbf{k}}^{\prime\prime} & -g_{m;\mathbf{k}}^{\prime}
\end{array}\right).
\label{eq:hq}
 \end{equation}
The energy eigenvalues are given by 
$\epsilon_{\mathbf{k}} = |f_{\mathbf{k}}| = 
\frac{J}{2}[3+2\cos k_1 + 2\cos k_2 + 2\cos (k_1-k_2)]^{1/2}$, with 
$f_{\mathbf{k}}= J \sum_{\bm{\delta}} t_{\bm{\delta};\mathbf{k}}$, and 
$t_{\bm{\delta};\mathbf{k}}= -\frac{i}{2}   e^{-i\mathbf{k}\cdot\bm{\delta}}$. 
The real ($g^{\prime}$) and imaginary part ($g^{\prime\prime}$) of the function 
$g$ of the scattering matrix $V$ are given by 
$g_{m;\mathbf{k}} =
(f_{\mathbf{k}}^*/\epsilon_{\mathbf{k}})
\sum_{\bm{\delta}} \Lambda_{m}^{\bm{\delta}} 
t_{\bm{\delta};\mathbf{k}} / \sqrt{N}$.  

The evaluation of the self energy using  \eqref{eq:SE} 
and \eqref{eq:hq} is straightforward. In that process and due to
$c_{\mathbf{k}}^{\dagger}=c_{-\mathbf{k}}$, anomalous commutators
$\{d_{1,\mathbf{k}},d_{2,\mathbf{k}'}\}=\delta_{\mathbf{k},-\mathbf{k}'}$
and their corresponding contractions arise, implying also a time evolution
$d_{j,\mathbf{k}}(t)=d_{j,\mathbf{k}}e^{\mp i\epsilon_{\mathbf{k}}t}$
for $j=1,2$, respectively. Thus, the self-energy exhibits particle-hole 
(ph) and particle-particle (pp) absorption channels, the amplitudes of 
which are determined via the diagonal and off-diagonal matrix elements of 
the matrix $V$. In the $q=0$ limit, considered here, the ph-channel vanishes.  
The pp-scattering-amplitudes acquire simple forms, and all diagonal 
and off-diagonal self-energies are 
described by a \emph{single function}
\begin{equation}\label{eq:Sigma}
\hspace{-3mm}
 \bm{\Sigma}(z) =
  \left(
  \begin{array}{cc}
    a_1 & b \\
    b  & a_2
  \end{array}
  \right) 
 \Sigma(z),
 \hphantom{a.}
 \Sigma(z) = \frac{1}{N} \sum_{\mathbf{k},\pm} 
 A_{\mathbf{k}}^2
 \frac{1-2f_{\mathbf{k}}}{2\epsilon_{\mathbf{k}} \mp z}, 
\end{equation}
with the matrix elements $a_1=\Lambda_1(3+\lambda_1^2)/\Lambda_2$, $a_2 =
\Lambda_2 (1-\lambda_2)^2 / \Lambda_1$, and $b=\lambda_1(1-\lambda_2)$, the
scattering amplitude $A_{\mathbf{k}} = \Lambda_1 \Lambda_2 \cos(k_xa/2)
\sin(\sqrt{3}k_ya/2) / (2 \epsilon_{\mathbf{k}})$, and the Fermi-Dirac
distribution $f_{\mathbf{k}}=1/(e^{\epsilon_{\mathbf{k}}/T}+1)$.  Two special
cases arise. For $\lambda_{2}=1$ the $E_{g_{2}}$ spin phonon Hamiltonian
satisfies $\mathcal{H}_{2}\propto H_{K}$, leading to no scattering for that
mode. For $\lambda_{1}=0$, symmetry prevents mixing of the $E_{g_{1}}$ and
$E_{g_{2}}$ modes.

\emph{Phonon self-energy: random gauge.--} For temperatures $T\gtrsim
0.1\dots0.2J \gg T^{*}$, thermal gauge excitations need to be taken into
account. For that, a random averaging over maximally disordered
configurations of $\eta_{\bm{\delta}_{3}}(\mathbf{r})$ is sufficient to
describe the fermionic system's properties. As this breaks translational
invariance, we resort to a numerical real-space evaluation of the
defect-averaged self-energy $\bm{\Sigma}(z)$.  This approach has been
detailed in Refs.~\cite{PhysRevB.101.035103,PhysRevB.96.205121} and is
recapitulated in \cite{supplement}.

\begin{figure}
\centering{}
\includegraphics[height=0.3\textheight]{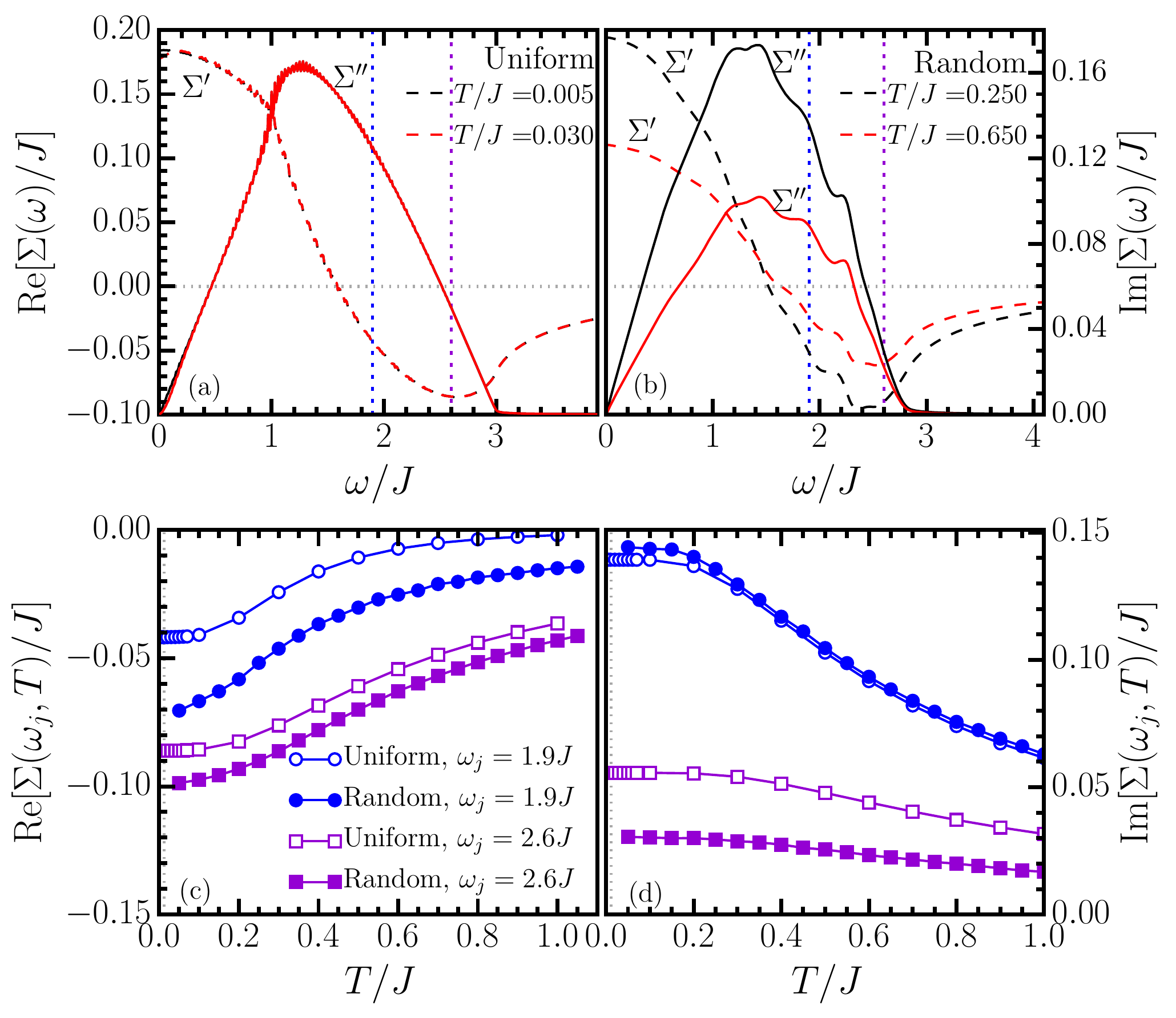}
\caption{
Panels (a) and (b): Frequency dependence of the self-energy $\Sigma$ 
for two temperatures in the (a) 
uniform gauge field sector for $L=200$ (curves overlap) and (b) 
averaged over $N_R=200$ maximally disordered 
sectors, on lattices of $L=30$, with $\Lambda_{1,2}/J=1$.   
Dashed blue(purple) lines at $\omega_j/J=1.9(2.6)$ on both panels indicate 
the position of the phonon modes $E_{g_{1}}$($E_{g_{2}}$). 
The scale of $\Sigma'$, for both uniform or random sectors, is indicated 
on the left side of panel (a) while the scale of $\Sigma''$ is shown on 
the right side of panel (b). 
Panels (c) and (d): Temperature  dependence of the real and imaginary 
parts of $\Sigma$, respectively,  at the phonon frequencies  
$\omega/J=1.9$ (blue circles) and $\omega/J=2.6$ (purple squares).  
Open(Filled) symbols indicate results acquired for uniform(random) 
gauge-field configuration(s).  
\label{fig:Sigma}} 
\end{figure}

\emph{Phonon self-energy: results.--} In Fig.~\ref{fig:Sigma}, we present
results for the self-energy $\Sigma$ versus frequency and temperature, for
both homogeneous as well as random gauge states. In Fig.~\ref{fig:Sigma}(a),
we plot $\Sigma(\omega)$ obtained from the analytical approach for two
temperatures, $T=0.005J<T^*$ and $T=0.030J \gtrsim T^*$ (curves overlap).
Calculations are performed
on lattices with $L=200$. For $\omega<0$, analyticity requires $(-)\Sigma^{
\prime ( \prime\prime )} (-\omega) = \Sigma^{ \prime (\prime\prime) }
(\omega)$ for the real(imaginary) part [left(right) $y$-axis of panel
[a(b)]]. Both, $\Sigma^\prime$ and $\Sigma^{\prime\prime}$, contribute to the
renormalization of the $E_{g_{1}}$ and $E_{g_{2}}$ phonons, the energy of
which is marked with a blue and purple dotted line, respectively. Since
$\Sigma'(\omega_{j},T)<0$ at both anticipated phonon energies $\omega_{j}$, a
downward renormalization will occur. Most importantly however, regarding
their life time, both phonons reside in a range of negative slope of
$\Sigma''$ versus $\omega$. Therefore the life time will be dispersive, with
phonon spectral functions that display asymmetric line shapes with enhanced
left-broadening.

Regarding the gauge excitations, Fig.~\ref{fig:Sigma}(b) shows $\Sigma$ from
the numerical approach for $T\gg T^\star$ on lattices of $L=30$, averaging
over $N_R=200$ maximally disordered gauge field configurations. First, the
overall shape of $\Sigma$ displays no qualitative change compared to
Fig.~\ref{fig:Sigma}(a), with only some additional fine structure arising
from the gauge-field excitations.  Second however, there is a reduction of
the bandwidth to $<3J$, which affects the scattering of the high frequency
mode $E_{g_{2}}$. Finally, there is clearly visible reduction of $\Sigma$
with increasing $T$. This is dictated by the Fermi-function (as in
Eq.~\eqref{eq:Sigma}) reducing the available phase space. For the low-$T$
results of Fig.~\ref{fig:Sigma}(a) this effect is too small to be observable.

In Figs.~\ref{fig:Sigma}(c) and (d), we scan the temperature dependence of
the real and imaginary part of the self-energy, respectively, at the two
frequencies of the phonon modes. In doing so, we plot results obtained from
both, the analytical and the numerical approach - even though in principle
the former is applied for temperatures beyond its validity. This figure
clearly demonstrates, that for the renormalization of optical phonons
gauge-field excitations yield only small quantitative corrections. For the
rest of the paper, we therefore remain in the uniform gauge configuration.

\begin{figure}
\centering{}
\includegraphics[height=0.3\textheight]{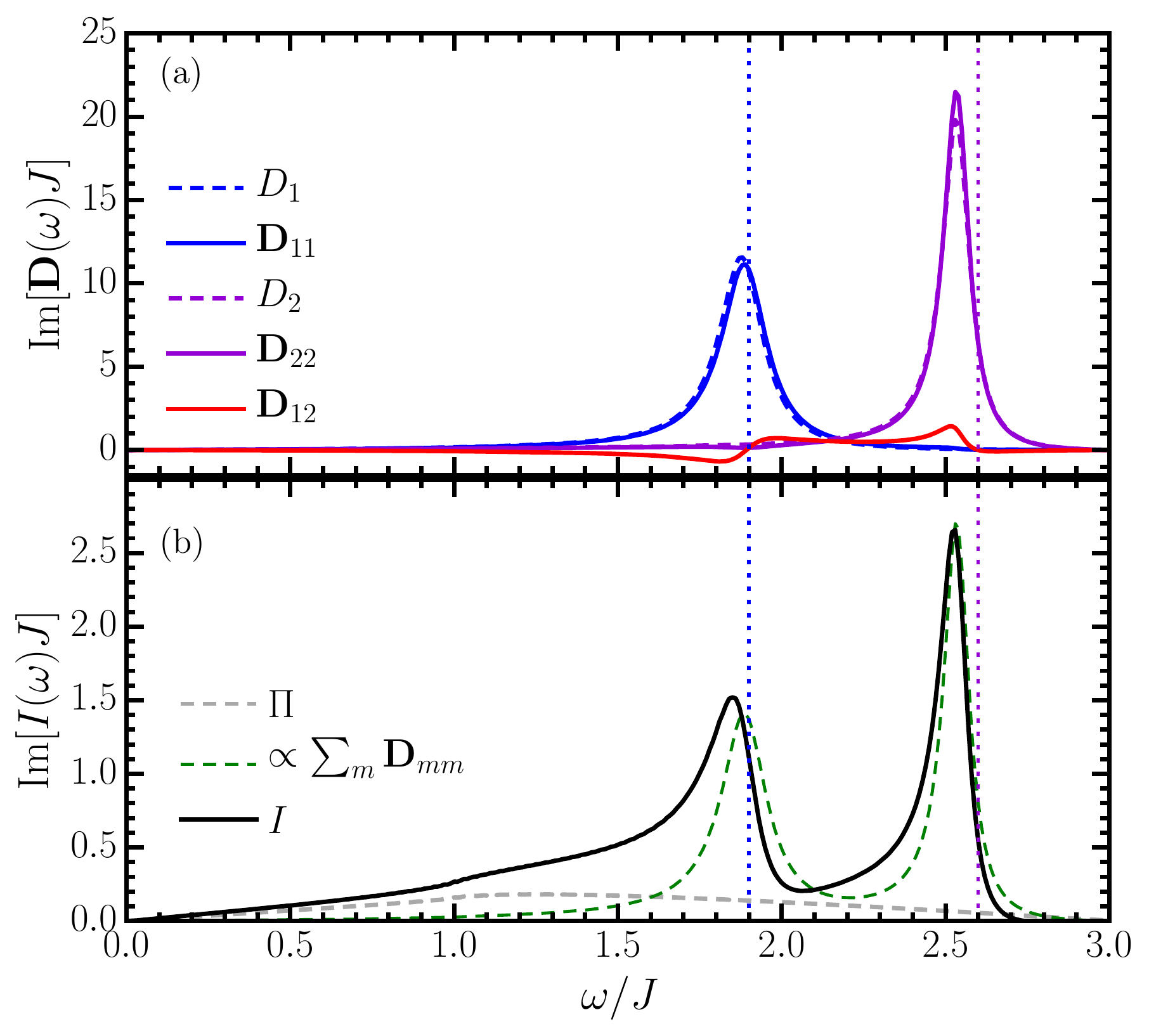}
\caption{
(a) Frequency dependence of the phonon spectral weight as evaluated via 
Eq.~\eqref{eq:SE} using Eq.~\eqref{eq:Sigma} for 
$\Lambda_{1(2)}/J = 0.3(0.9)$, $\lambda_{1,2}=2$, and  $T=0.005J$.   
Dashed lines represent the decoupled phonon modes $D_{m}$.  
(b) Frequency dependence of the total Raman intensity $I$ (black solid line), 
evaluated via Eq.~\eqref{eq:Raman} for the same $\bm{\Lambda}$'s as for panel (a) 
and $r_{m}=0.3$, $C_{m}=-0.8J$. For comparison, the diagonal  
phonon spectral weight 
$\sum_{m} \mathbf{D}_{mm}(\omega)$ is plotted (green dashed line), 
here rescaled by a factor of $1/8$ for visual reasons, 
as well as the fermionic Raman response, 
$\sim \Pi(\omega)$ (gray dashed line). 
\label{fig:Fano}} 
\end{figure}

Next we discuss the renormalized optical phonon modes, obtained from the
$2\times2$ Dyson equation Eq.~\eqref{eq:SE} using $\bm{\Sigma}$ from
Eq.~\eqref{eq:Sigma}. In general, because of $b\neq 0$, phonon mixing by
virtue of the fermionic background will occur, i.e. $\mathbf{D}_{mm} \neq
[D_{m}^0 - a_m \Sigma]^{-1}\equiv D_m$. However, we will stay in the perturbative
regime $b\Sigma((\omega_2+\omega_1)/2)/(\omega_2-\omega_1)\ll 1$, where the
mixing is weak and can be ignored. Fig.~\ref{fig:Fano}(a) depicts the
individual (off)diagonal phonon spectra $D''_{mn}(\omega)$ for
intermediate $\Lambda_{1}(\Lambda_2) =0.3J(0.9J)$, $\lambda_{1,2}=2$ and
for $T=0.005J$.  First, both diagonal elements of $\mathbf{D}$ exhibit the
anticipated downward renormalization with respect to $\omega_m$, largest for
$E_{g_2}$. In accordance with Fig.~\ref{fig:Sigma}(c), the modes' peaks will
shift upwards to $\omega_m$ as the temperature is increased. Second, the
frequency dependence of $\Sigma''(\omega)$ leads to slightly asymmetric phonon
line shapes. We emphasize, that this asymmetry is {\em not} related to the Fano
effect \cite{PhysRev.124.1866}, commonly cited in such cases. The width of
the phonon mode is expected to shrink as the temperature is increased,
Fig.~\ref{fig:Sigma}(d).  I.e., in contrast to most conventional excitations
in many-body systems, their life-time grows with temperature. Third, the
phonon mixing is rather weak, i.e., $\mathbf{D}_{12} \ll \mathbf{D}_{11(22)}$
and $\mathbf{D}_{mm} \approx D_m $, essentially rendering the two phonon
modes decoupled.

\emph{Raman response and Fano-lineshape.--} Finally, we speculate on the Raman cross section $I(z)$. Light scatters from
both, the lattice \emph{and} the magnetic degrees of freedom. {\em This}
forces the Fano effect to occur \cite{PhysRev.124.1866} and renders the
Raman cross section a coupled three-channel problem, with Raman vertices $F$
and $R_{m=1,2}$, encoding the couplings of incoming(outgoing) light fields to
the fermions and the two phonon modes, respectively. Presently, only the
Loudon-Fleury vertex $F$ is known microscopically \cite{PhysRev.166.514,
PhysRevLett.113.187201, Nasu2016}. Therefore, obtaining $I(z)$ from first
principles is infeasible. To make progress, we resort to
\emph{phenomenological} simplifications. These are detailed in the
supplemental material \cite{supplement}, but essentially amount to: (i) We
replace the Raman vertices $F$ and $R_{m=1,2}$ by mere constants, dependent
on the scattering geometry, (ii) We approximate all fermionic two-particle
Greens functions of the three-channel problem beyond $\Sigma(\omega)$ by the
latter. Contrasting Figs.~\ref{fig:Sigma}(a,b) against the known {\em
magnetic} Raman response \cite{PhysRevLett.113.187201, Nasu2016} this is
acceptable. (iii) We ignore phonon-mixing. This leads to \cite{supplement}
\begin{equation}
I(z)\approx \Pi(z)+\sum_{m=1,2}[r_{m}+C_{m}\Pi(z)]^{2}D_{mm}(z)\,,
\label{eq:Raman}
\end{equation}
where $\Pi(z)=\Sigma(z)[J/(\Lambda_1\Lambda_2)]^2$ and $I(z)$ is here normalized
to $F$, i.e., $I(z) \rightarrow I(z)/F^2$, leaving four free parameters, namely
$r_{m}=R_{m}/F$ and the coupling constants $C_{m}$, which allows to adjust
the strength of the Fano effect. $I(z)$ is a retarded propagator and maps
to the cross section by the fluctuation-dissipation
prefactor. 

In Fig.~\ref{fig:Fano}(b), we plot $I''(\omega)$, 
for empirically chosen $r_{m}=0.3$, $C_{m}=-0.8J$, 
together with $\Pi''(\omega)$, 
as well as the rescaled diagonal intensity $\sum_{m}\mathbf{D}''_{mm}(\omega)$
from Fig.~\ref{fig:Fano}(a). The spectrum is remarkably similar to the
experimental findings in Refs.~\cite{PhysRevLett.114.147201,
PhysRevB.95.174429, Wulferding2020, PhysRevB.101.045419,
PhysRevB.100.134419}.  A broad continuum due to the 
fractionalized magnetic excitations is visible, on top of which two asymmetric 
phonon line-shapes ride, with a characteristic sharp, almost vertical, drop-off 
only on the high frequency side of the mode.  The overall shape of the 
Raman response $I''(\omega)$ is distinctly different from the diagonal phonon 
intensity, $ \propto \sum_{m}\mathbf{D}''_{mm}(\omega)$, green dashed line in 
Fig.~\ref{fig:Fano}(b), which would approximately describe
the response, if only the Majorana-Phonon scattering was taken  
into account but not the Fano effect. 

\emph{Conclusion.--}
We have provided a microscopic theory of optical phonons coupled to a KSL. Our analysis strongly supports the origin of the Fano
line-shapes observed in Raman experiments on Kitaev candidate materials to be
due to the fractionalized magnetic excitations. Moreover, direct comparison
between Fig.~\ref{fig:Fano}(a) and (b) shows that the origin of the
line-shape asymmetry is twofold, namely stemming from the Majorana-phonon
scattering itself, as well as from the Fano effect.

In the future, it would be desirable to obtain the spin-phonon coupling constants from ab-initio calculations \cite{kaib2020magnetoelastic} for a quantitative description and to extend the analysis to other QSL candidate materials. We also expect that a better understanding  spin-phonon coupling is crucial for understanding the thermal transport behavior of $\alpha$-RuCl$_3$.   

\emph{Acknowledgements.--} We acknowledge helpful discussions with
D. Wulferding, K. Burch, P. Lemmens, S. Bhattacharjee and R. Moessner. We
are grateful to R. Valent\'{i} and S. Biswas for clarifications on optical
phonons, extracted from ab-initio methods. Work of A.M. and W.B. has been
supported in part by the DFG through Project A02 of SFB 1143 (Project-Id
247310070), by Nds.~QUANOMET, and by the National Science Foundation under
Grant No.~NSF PHY-1748958.  W.B.~also acknowledges the kind hospitality of
the PSM, Dresden.

\begin{thebibliography}{61}%
\makeatletter
\providecommand \@ifxundefined [1]{%
 \@ifx{#1\undefined}
}%
\providecommand \@ifnum [1]{%
 \ifnum #1\expandafter \@firstoftwo
 \else \expandafter \@secondoftwo
 \fi
}%
\providecommand \@ifx [1]{%
 \ifx #1\expandafter \@firstoftwo
 \else \expandafter \@secondoftwo
 \fi
}%
\providecommand \natexlab [1]{#1}%
\providecommand \enquote  [1]{``#1''}%
\providecommand \bibnamefont  [1]{#1}%
\providecommand \bibfnamefont [1]{#1}%
\providecommand \citenamefont [1]{#1}%
\providecommand \href@noop [0]{\@secondoftwo}%
\providecommand \href [0]{\begingroup \@sanitize@url \@href}%
\providecommand \@href[1]{\@@startlink{#1}\@@href}%
\providecommand \@@href[1]{\endgroup#1\@@endlink}%
\providecommand \@sanitize@url [0]{\catcode `\\12\catcode `\$12\catcode
  `\&12\catcode `\#12\catcode `\^12\catcode `\_12\catcode `\%12\relax}%
\providecommand \@@startlink[1]{}%
\providecommand \@@endlink[0]{}%
\providecommand \url  [0]{\begingroup\@sanitize@url \@url }%
\providecommand \@url [1]{\endgroup\@href {#1}{\urlprefix }}%
\providecommand \urlprefix  [0]{URL }%
\providecommand \Eprint [0]{\href }%
\providecommand \doibase [0]{https://doi.org/}%
\providecommand \selectlanguage [0]{\@gobble}%
\providecommand \bibinfo  [0]{\@secondoftwo}%
\providecommand \bibfield  [0]{\@secondoftwo}%
\providecommand \translation [1]{[#1]}%
\providecommand \BibitemOpen [0]{}%
\providecommand \bibitemStop [0]{}%
\providecommand \bibitemNoStop [0]{.\EOS\space}%
\providecommand \EOS [0]{\spacefactor3000\relax}%
\providecommand \BibitemShut  [1]{\csname bibitem#1\endcsname}%
\let\auto@bib@innerbib\@empty
\bibitem [{\citenamefont {Sandilands}\ \emph {et~al.}(2015)\citenamefont
  {Sandilands}, \citenamefont {Tian}, \citenamefont {Plumb}, \citenamefont
  {Kim},\ and\ \citenamefont {Burch}}]{PhysRevLett.114.147201}%
  \BibitemOpen
  \bibfield  {author} {\bibinfo {author} {\bibfnamefont {L.~J.}\ \bibnamefont
  {Sandilands}}, \bibinfo {author} {\bibfnamefont {Y.}~\bibnamefont {Tian}},
  \bibinfo {author} {\bibfnamefont {K.~W.}\ \bibnamefont {Plumb}}, \bibinfo
  {author} {\bibfnamefont {Y.-J.}\ \bibnamefont {Kim}},\ and\ \bibinfo {author}
  {\bibfnamefont {K.~S.}\ \bibnamefont {Burch}},\ }\href
  {https://doi.org/10.1103/PhysRevLett.114.147201} {\bibfield  {journal}
  {\bibinfo  {journal} {Phys. Rev. Lett.}\ }\textbf {\bibinfo {volume} {114}},\
  \bibinfo {pages} {147201} (\bibinfo {year} {2015})}\BibitemShut {NoStop}%
\bibitem [{\citenamefont {Glamazda}\ \emph {et~al.}(2017)\citenamefont
  {Glamazda}, \citenamefont {Lemmens}, \citenamefont {Do}, \citenamefont
  {Kwon},\ and\ \citenamefont {Choi}}]{PhysRevB.95.174429}%
  \BibitemOpen
  \bibfield  {author} {\bibinfo {author} {\bibfnamefont {A.}~\bibnamefont
  {Glamazda}}, \bibinfo {author} {\bibfnamefont {P.}~\bibnamefont {Lemmens}},
  \bibinfo {author} {\bibfnamefont {S.-H.}\ \bibnamefont {Do}}, \bibinfo
  {author} {\bibfnamefont {Y.~S.}\ \bibnamefont {Kwon}},\ and\ \bibinfo
  {author} {\bibfnamefont {K.-Y.}\ \bibnamefont {Choi}},\ }\href
  {https://doi.org/10.1103/PhysRevB.95.174429} {\bibfield  {journal} {\bibinfo
  {journal} {Phys. Rev. B}\ }\textbf {\bibinfo {volume} {95}},\ \bibinfo
  {pages} {174429} (\bibinfo {year} {2017})}\BibitemShut {NoStop}%
\bibitem [{\citenamefont {Mai}\ \emph {et~al.}(2019)\citenamefont {Mai},
  \citenamefont {McCreary}, \citenamefont {Lampen-Kelley}, \citenamefont
  {Butch}, \citenamefont {Simpson}, \citenamefont {Yan}, \citenamefont
  {Nagler}, \citenamefont {Mandrus}, \citenamefont {Walker},\ and\
  \citenamefont {Aguilar}}]{PhysRevB.100.134419}%
  \BibitemOpen
  \bibfield  {author} {\bibinfo {author} {\bibfnamefont {T.~T.}\ \bibnamefont
  {Mai}}, \bibinfo {author} {\bibfnamefont {A.}~\bibnamefont {McCreary}},
  \bibinfo {author} {\bibfnamefont {P.}~\bibnamefont {Lampen-Kelley}}, \bibinfo
  {author} {\bibfnamefont {N.}~\bibnamefont {Butch}}, \bibinfo {author}
  {\bibfnamefont {J.~R.}\ \bibnamefont {Simpson}}, \bibinfo {author}
  {\bibfnamefont {J.-Q.}\ \bibnamefont {Yan}}, \bibinfo {author} {\bibfnamefont
  {S.~E.}\ \bibnamefont {Nagler}}, \bibinfo {author} {\bibfnamefont
  {D.}~\bibnamefont {Mandrus}}, \bibinfo {author} {\bibfnamefont {A.~R.~H.}\
  \bibnamefont {Walker}},\ and\ \bibinfo {author} {\bibfnamefont {R.~V.}\
  \bibnamefont {Aguilar}},\ }\href
  {https://doi.org/10.1103/PhysRevB.100.134419} {\bibfield  {journal} {\bibinfo
   {journal} {Phys. Rev. B}\ }\textbf {\bibinfo {volume} {100}},\ \bibinfo
  {pages} {134419} (\bibinfo {year} {2019})}\BibitemShut {NoStop}%
\bibitem [{\citenamefont {Lin}\ \emph {et~al.}(2020)\citenamefont {Lin},
  \citenamefont {Ran}, \citenamefont {Zheng}, \citenamefont {Xu}, \citenamefont
  {Gao}, \citenamefont {Wen}, \citenamefont {Yu}, \citenamefont {Li},\ and\
  \citenamefont {Xi}}]{PhysRevB.101.045419}%
  \BibitemOpen
  \bibfield  {author} {\bibinfo {author} {\bibfnamefont {D.}~\bibnamefont
  {Lin}}, \bibinfo {author} {\bibfnamefont {K.}~\bibnamefont {Ran}}, \bibinfo
  {author} {\bibfnamefont {H.}~\bibnamefont {Zheng}}, \bibinfo {author}
  {\bibfnamefont {J.}~\bibnamefont {Xu}}, \bibinfo {author} {\bibfnamefont
  {L.}~\bibnamefont {Gao}}, \bibinfo {author} {\bibfnamefont {J.}~\bibnamefont
  {Wen}}, \bibinfo {author} {\bibfnamefont {S.-L.}\ \bibnamefont {Yu}},
  \bibinfo {author} {\bibfnamefont {J.-X.}\ \bibnamefont {Li}},\ and\ \bibinfo
  {author} {\bibfnamefont {X.}~\bibnamefont {Xi}},\ }\href@noop {} {\bibfield
  {journal} {\bibinfo  {journal} {Phys. Rev. B}\ }\textbf {\bibinfo {volume}
  {101}},\ \bibinfo {pages} {045419} (\bibinfo {year} {2020})}\BibitemShut
  {NoStop}%
\bibitem [{\citenamefont {Wulferding}\ \emph {et~al.}(2020)\citenamefont
  {Wulferding}, \citenamefont {Choi}, \citenamefont {Do}, \citenamefont {Lee},
  \citenamefont {Lemmens}, \citenamefont {Faugeras}, \citenamefont {Gallais},\
  and\ \citenamefont {Choi}}]{Wulferding2020}%
  \BibitemOpen
  \bibfield  {author} {\bibinfo {author} {\bibfnamefont {D.}~\bibnamefont
  {Wulferding}}, \bibinfo {author} {\bibfnamefont {Y.}~\bibnamefont {Choi}},
  \bibinfo {author} {\bibfnamefont {S.-H.}\ \bibnamefont {Do}}, \bibinfo
  {author} {\bibfnamefont {C.~H.}\ \bibnamefont {Lee}}, \bibinfo {author}
  {\bibfnamefont {P.}~\bibnamefont {Lemmens}}, \bibinfo {author} {\bibfnamefont
  {C.}~\bibnamefont {Faugeras}}, \bibinfo {author} {\bibfnamefont
  {Y.}~\bibnamefont {Gallais}},\ and\ \bibinfo {author} {\bibfnamefont {K.-Y.}\
  \bibnamefont {Choi}},\ }\href@noop {} {\bibfield  {journal} {\bibinfo
  {journal} {Nature Communications}\ }\textbf {\bibinfo {volume} {11}},\
  \bibinfo {pages} {1603} (\bibinfo {year} {2020})}\BibitemShut {NoStop}%
\bibitem [{\citenamefont {Wang}\ \emph {et~al.}(2020)\citenamefont {Wang},
  \citenamefont {Osterhoudt}, \citenamefont {Tian}, \citenamefont
  {Lampen-Kelley}, \citenamefont {Banerjee}, \citenamefont {Goldstein},
  \citenamefont {Yan}, \citenamefont {Knolle}, \citenamefont {Ji},
  \citenamefont {Cava} \emph {et~al.}}]{wang2020range}%
  \BibitemOpen
  \bibfield  {author} {\bibinfo {author} {\bibfnamefont {Y.}~\bibnamefont
  {Wang}}, \bibinfo {author} {\bibfnamefont {G.~B.}\ \bibnamefont
  {Osterhoudt}}, \bibinfo {author} {\bibfnamefont {Y.}~\bibnamefont {Tian}},
  \bibinfo {author} {\bibfnamefont {P.}~\bibnamefont {Lampen-Kelley}}, \bibinfo
  {author} {\bibfnamefont {A.}~\bibnamefont {Banerjee}}, \bibinfo {author}
  {\bibfnamefont {T.}~\bibnamefont {Goldstein}}, \bibinfo {author}
  {\bibfnamefont {J.}~\bibnamefont {Yan}}, \bibinfo {author} {\bibfnamefont
  {J.}~\bibnamefont {Knolle}}, \bibinfo {author} {\bibfnamefont
  {H.}~\bibnamefont {Ji}}, \bibinfo {author} {\bibfnamefont {R.~J.}\
  \bibnamefont {Cava}}, \emph {et~al.},\ }\href@noop {} {\bibfield  {journal}
  {\bibinfo  {journal} {npj Quantum Materials}\ }\textbf {\bibinfo {volume}
  {5}},\ \bibinfo {pages} {1} (\bibinfo {year} {2020})}\BibitemShut {NoStop}%
\bibitem [{\citenamefont {Savary}\ and\ \citenamefont
  {Balents}(2017)}]{0034-4885-80-1-016502}%
  \BibitemOpen
  \bibfield  {author} {\bibinfo {author} {\bibfnamefont {L.}~\bibnamefont
  {Savary}}\ and\ \bibinfo {author} {\bibfnamefont {L.}~\bibnamefont
  {Balents}},\ }\href@noop {} {\bibfield  {journal} {\bibinfo  {journal}
  {Reports on Progress in Physics}\ }\textbf {\bibinfo {volume} {80}},\
  \bibinfo {pages} {016502} (\bibinfo {year} {2017})}\BibitemShut {NoStop}%
\bibitem [{\citenamefont {Zhou}\ \emph {et~al.}(2017)\citenamefont {Zhou},
  \citenamefont {Kanoda},\ and\ \citenamefont {Ng}}]{RevModPhys.89.025003}%
  \BibitemOpen
  \bibfield  {author} {\bibinfo {author} {\bibfnamefont {Y.}~\bibnamefont
  {Zhou}}, \bibinfo {author} {\bibfnamefont {K.}~\bibnamefont {Kanoda}},\ and\
  \bibinfo {author} {\bibfnamefont {T.-K.}\ \bibnamefont {Ng}},\ }\href@noop {}
  {\bibfield  {journal} {\bibinfo  {journal} {Rev. Mod. Phys.}\ }\textbf
  {\bibinfo {volume} {89}},\ \bibinfo {pages} {025003} (\bibinfo {year}
  {2017})}\BibitemShut {NoStop}%
\bibitem [{\citenamefont {Knolle}\ and\ \citenamefont
  {Moessner}(2019)}]{doi:10.1146/annurev-conmatphys-031218-013401}%
  \BibitemOpen
  \bibfield  {author} {\bibinfo {author} {\bibfnamefont {J.}~\bibnamefont
  {Knolle}}\ and\ \bibinfo {author} {\bibfnamefont {R.}~\bibnamefont
  {Moessner}},\ }\href@noop {} {\bibfield  {journal} {\bibinfo  {journal}
  {Annual Review of Condensed Matter Physics}\ }\textbf {\bibinfo {volume}
  {10}},\ \bibinfo {pages} {451} (\bibinfo {year} {2019})}\BibitemShut
  {NoStop}%
\bibitem [{\citenamefont {Kitaev}(2006)}]{Kitaev20062}%
  \BibitemOpen
  \bibfield  {author} {\bibinfo {author} {\bibfnamefont {A.}~\bibnamefont
  {Kitaev}},\ }\href@noop {} {\bibfield  {journal} {\bibinfo  {journal} {Annals
  of Physics}\ }\textbf {\bibinfo {volume} {321}},\ \bibinfo {pages} {2}
  (\bibinfo {year} {2006})},\ \bibinfo {note} {january Special
  Issue}\BibitemShut {NoStop}%
\bibitem [{\citenamefont {Steinigeweg}\ and\ \citenamefont
  {Brenig}(2016)}]{PhysRevB.93.214425}%
  \BibitemOpen
  \bibfield  {author} {\bibinfo {author} {\bibfnamefont {R.}~\bibnamefont
  {Steinigeweg}}\ and\ \bibinfo {author} {\bibfnamefont {W.}~\bibnamefont
  {Brenig}},\ }\href@noop {} {\bibfield  {journal} {\bibinfo  {journal} {Phys.
  Rev. B}\ }\textbf {\bibinfo {volume} {93}},\ \bibinfo {pages} {214425}
  (\bibinfo {year} {2016})}\BibitemShut {NoStop}%
\bibitem [{\citenamefont {Feng}\ \emph {et~al.}(2007)\citenamefont {Feng},
  \citenamefont {Zhang},\ and\ \citenamefont {Xiang}}]{PhysRevLett.98.087204}%
  \BibitemOpen
  \bibfield  {author} {\bibinfo {author} {\bibfnamefont {X.-Y.}\ \bibnamefont
  {Feng}}, \bibinfo {author} {\bibfnamefont {G.-M.}\ \bibnamefont {Zhang}},\
  and\ \bibinfo {author} {\bibfnamefont {T.}~\bibnamefont {Xiang}},\
  }\href@noop {} {\bibfield  {journal} {\bibinfo  {journal} {Phys. Rev. Lett.}\
  }\textbf {\bibinfo {volume} {98}},\ \bibinfo {pages} {087204} (\bibinfo
  {year} {2007})}\BibitemShut {NoStop}%
\bibitem [{\citenamefont {Wu}(2012)}]{Wu20123530}%
  \BibitemOpen
  \bibfield  {author} {\bibinfo {author} {\bibfnamefont {N.}~\bibnamefont
  {Wu}},\ }\href@noop {} {\bibfield  {journal} {\bibinfo  {journal} {Physics
  Letters A}\ }\textbf {\bibinfo {volume} {376}},\ \bibinfo {pages} {3530}
  (\bibinfo {year} {2012})}\BibitemShut {NoStop}%
\bibitem [{\citenamefont {Metavitsiadis}\ and\ \citenamefont
  {Brenig}(2017)}]{PhysRevB.96.041115}%
  \BibitemOpen
  \bibfield  {author} {\bibinfo {author} {\bibfnamefont {A.}~\bibnamefont
  {Metavitsiadis}}\ and\ \bibinfo {author} {\bibfnamefont {W.}~\bibnamefont
  {Brenig}},\ }\href@noop {} {\bibfield  {journal} {\bibinfo  {journal} {Phys.
  Rev. B}\ }\textbf {\bibinfo {volume} {96}},\ \bibinfo {pages} {041115}
  (\bibinfo {year} {2017})}\BibitemShut {NoStop}%
\bibitem [{\citenamefont {Metavitsiadis}\ \emph {et~al.}(2019)\citenamefont
  {Metavitsiadis}, \citenamefont {Psaroudaki},\ and\ \citenamefont
  {Brenig}}]{PhysRevB.99.205129}%
  \BibitemOpen
  \bibfield  {author} {\bibinfo {author} {\bibfnamefont {A.}~\bibnamefont
  {Metavitsiadis}}, \bibinfo {author} {\bibfnamefont {C.}~\bibnamefont
  {Psaroudaki}},\ and\ \bibinfo {author} {\bibfnamefont {W.}~\bibnamefont
  {Brenig}},\ }\href@noop {} {\bibfield  {journal} {\bibinfo  {journal} {Phys.
  Rev. B}\ }\textbf {\bibinfo {volume} {99}},\ \bibinfo {pages} {205129}
  (\bibinfo {year} {2019})}\BibitemShut {NoStop}%
\bibitem [{\citenamefont {Agrapidis}\ \emph {et~al.}(2019)\citenamefont
  {Agrapidis}, \citenamefont {van~den Brink},\ and\ \citenamefont
  {Nishimoto}}]{PhysRevB.99.224418}%
  \BibitemOpen
  \bibfield  {author} {\bibinfo {author} {\bibfnamefont {C.~E.}\ \bibnamefont
  {Agrapidis}}, \bibinfo {author} {\bibfnamefont {J.}~\bibnamefont {van~den
  Brink}},\ and\ \bibinfo {author} {\bibfnamefont {S.}~\bibnamefont
  {Nishimoto}},\ }\href@noop {} {\bibfield  {journal} {\bibinfo  {journal}
  {Phys. Rev. B}\ }\textbf {\bibinfo {volume} {99}},\ \bibinfo {pages} {224418}
  (\bibinfo {year} {2019})}\BibitemShut {NoStop}%
\bibitem [{\citenamefont {Metavitsiadis}\ and\ \citenamefont
  {Brenig}(2020{\natexlab{a}})}]{metavitsiadis2020flux}%
  \BibitemOpen
  \bibfield  {author} {\bibinfo {author} {\bibfnamefont {A.}~\bibnamefont
  {Metavitsiadis}}\ and\ \bibinfo {author} {\bibfnamefont {W.}~\bibnamefont
  {Brenig}},\ }\href@noop {} {} (\bibinfo {year} {2020}{\natexlab{a}}),\
  \Eprint {https://arxiv.org/abs/2009.04467} {arXiv:2009.04467
  [cond-mat.str-el]} \BibitemShut {NoStop}%
\bibitem [{Yan()}]{Yang2007}%
  \BibitemOpen
  \href@noop {} {\bibinfo  {journal} {S. Yang, D. L. Zhou, and C. P. Sun, Phys.
  Rev. B {\bf 76}, 180404 (2007)}\ }\BibitemShut {NoStop}%
\bibitem [{\citenamefont {Nasu}\ \emph {et~al.}(2014)\citenamefont {Nasu},
  \citenamefont {Udagawa},\ and\ \citenamefont
  {Motome}}]{PhysRevLett.113.197205}%
  \BibitemOpen
\bibfield  {journal} {  }\bibfield  {author} {\bibinfo {author} {\bibfnamefont
  {J.}~\bibnamefont {Nasu}}, \bibinfo {author} {\bibfnamefont {M.}~\bibnamefont
  {Udagawa}},\ and\ \bibinfo {author} {\bibfnamefont {Y.}~\bibnamefont
  {Motome}},\ }\href@noop {} {\bibfield  {journal} {\bibinfo  {journal} {Phys.
  Rev. Lett.}\ }\textbf {\bibinfo {volume} {113}},\ \bibinfo {pages} {197205}
  (\bibinfo {year} {2014})}\BibitemShut {NoStop}%
\bibitem [{OBr()}]{OBrien2016}%
  \BibitemOpen
  \href@noop {} {\bibinfo  {journal} {K. O'Brien, M. Hermanns, and S. Trebst,
  Phys. Rev. B {\bf 93}, 085101 (2016)}\ }\BibitemShut {NoStop}%
\bibitem [{\citenamefont {Mishchenko}\ \emph {et~al.}(2017)\citenamefont
  {Mishchenko}, \citenamefont {Kato},\ and\ \citenamefont
  {Motome}}]{PhysRevB.96.125124}%
  \BibitemOpen
\bibfield  {journal} {  }\bibfield  {author} {\bibinfo {author} {\bibfnamefont
  {P.~A.}\ \bibnamefont {Mishchenko}}, \bibinfo {author} {\bibfnamefont
  {Y.}~\bibnamefont {Kato}},\ and\ \bibinfo {author} {\bibfnamefont
  {Y.}~\bibnamefont {Motome}},\ }\href@noop {} {\bibfield  {journal} {\bibinfo
  {journal} {Phys. Rev. B}\ }\textbf {\bibinfo {volume} {96}},\ \bibinfo
  {pages} {125124} (\bibinfo {year} {2017})}\BibitemShut {NoStop}%
\bibitem [{\citenamefont {Baskaran}\ \emph {et~al.}(2008)\citenamefont
  {Baskaran}, \citenamefont {Sen},\ and\ \citenamefont
  {Shankar}}]{PhysRevB.78.115116}%
  \BibitemOpen
  \bibfield  {author} {\bibinfo {author} {\bibfnamefont {G.}~\bibnamefont
  {Baskaran}}, \bibinfo {author} {\bibfnamefont {D.}~\bibnamefont {Sen}},\ and\
  \bibinfo {author} {\bibfnamefont {R.}~\bibnamefont {Shankar}},\ }\href@noop
  {} {\bibfield  {journal} {\bibinfo  {journal} {Phys. Rev. B}\ }\textbf
  {\bibinfo {volume} {78}},\ \bibinfo {pages} {115116} (\bibinfo {year}
  {2008})}\BibitemShut {NoStop}%
\bibitem [{\citenamefont {Rousochatzakis}\ \emph {et~al.}(2018)\citenamefont
  {Rousochatzakis}, \citenamefont {Sizyuk},\ and\ \citenamefont
  {Perkins}}]{Rousochatzakis2018}%
  \BibitemOpen
  \bibfield  {author} {\bibinfo {author} {\bibfnamefont {I.}~\bibnamefont
  {Rousochatzakis}}, \bibinfo {author} {\bibfnamefont {Y.}~\bibnamefont
  {Sizyuk}},\ and\ \bibinfo {author} {\bibfnamefont {N.~B.}\ \bibnamefont
  {Perkins}},\ }\href@noop {} {\bibfield  {journal} {\bibinfo  {journal}
  {Nature Communications}\ }\textbf {\bibinfo {volume} {9}},\ \bibinfo {pages}
  {1575} (\bibinfo {year} {2018})}\BibitemShut {NoStop}%
\bibitem [{\citenamefont {Stavropoulos}\ \emph {et~al.}(2019)\citenamefont
  {Stavropoulos}, \citenamefont {Pereira},\ and\ \citenamefont
  {Kee}}]{PhysRevLett.123.037203}%
  \BibitemOpen
  \bibfield  {author} {\bibinfo {author} {\bibfnamefont {P.~P.}\ \bibnamefont
  {Stavropoulos}}, \bibinfo {author} {\bibfnamefont {D.}~\bibnamefont
  {Pereira}},\ and\ \bibinfo {author} {\bibfnamefont {H.-Y.}\ \bibnamefont
  {Kee}},\ }\href@noop {} {\bibfield  {journal} {\bibinfo  {journal} {Phys.
  Rev. Lett.}\ }\textbf {\bibinfo {volume} {123}},\ \bibinfo {pages} {037203}
  (\bibinfo {year} {2019})}\BibitemShut {NoStop}%
\bibitem [{Kha()}]{Khaliullin2005}%
  \BibitemOpen
  \href@noop {} {\bibinfo  {journal} {G. Khaliullin, Prog. Theor. Phys. Suppl.
  \textbf{160}, 155 (2005)}\ }\BibitemShut {NoStop}%
\bibitem [{\citenamefont {Jackeli}\ and\ \citenamefont
  {Khaliullin}(2009)}]{PhysRevLett.102.017205}%
  \BibitemOpen
\bibfield  {journal} {  }\bibfield  {author} {\bibinfo {author} {\bibfnamefont
  {G.}~\bibnamefont {Jackeli}}\ and\ \bibinfo {author} {\bibfnamefont
  {G.}~\bibnamefont {Khaliullin}},\ }\href
  {https://link.aps.org/doi/10.1103/PhysRevLett.102.017205} {\bibfield
  {journal} {\bibinfo  {journal} {Phys. Rev. Lett.}\ }\textbf {\bibinfo
  {volume} {102}},\ \bibinfo {pages} {017205} (\bibinfo {year}
  {2009})}\BibitemShut {NoStop}%
\bibitem [{\citenamefont {Chaloupka}\ \emph {et~al.}(2010)\citenamefont
  {Chaloupka}, \citenamefont {Jackeli},\ and\ \citenamefont
  {Khaliullin}}]{PhysRevLett.105.027204}%
  \BibitemOpen
  \bibfield  {author} {\bibinfo {author} {\bibfnamefont {J.}~\bibnamefont
  {Chaloupka}}, \bibinfo {author} {\bibfnamefont {G.}~\bibnamefont {Jackeli}},\
  and\ \bibinfo {author} {\bibfnamefont {G.}~\bibnamefont {Khaliullin}},\
  }\href@noop {} {\bibfield  {journal} {\bibinfo  {journal} {Phys. Rev. Lett.}\
  }\textbf {\bibinfo {volume} {105}},\ \bibinfo {pages} {027204} (\bibinfo
  {year} {2010})}\BibitemShut {NoStop}%
\bibitem [{\citenamefont {Nussinov}\ and\ \citenamefont {van~den
  Brink}(2015)}]{RevModPhys.87.1}%
  \BibitemOpen
  \bibfield  {author} {\bibinfo {author} {\bibfnamefont {Z.}~\bibnamefont
  {Nussinov}}\ and\ \bibinfo {author} {\bibfnamefont {J.}~\bibnamefont {van~den
  Brink}},\ }\href@noop {} {\bibfield  {journal} {\bibinfo  {journal} {Rev.
  Mod. Phys.}\ }\textbf {\bibinfo {volume} {87}},\ \bibinfo {pages} {1}
  (\bibinfo {year} {2015})}\BibitemShut {NoStop}%
\bibitem [{Tre()}]{Trebst2017}%
  \BibitemOpen
  \href@noop {} {\bibinfo  {journal} {S. Trebst, \emph{Kitaev Materials},
  Lecture Notes of the 48th IFF Spring School 2017, S. Bl{\"u}gel, Y.
  Mokrousov, T. Sch{\"a}pers, Y. Ando (Eds.), ISBN 978-3-95806-202-3}\
  }\BibitemShut {NoStop}%
\bibitem [{Win()}]{Winter2017}%
  \BibitemOpen
\bibfield  {journal} {  }\href@noop {} {\bibinfo  {journal} {S. M. Winter, A.
  A. Tsirlin, M. Daghofer, J. van den Brink, Y. Singh, P. Gegenwart, and R.
  Valent\'{\i}, J. Phys.: Condens. Matter {\bf 29}, 493002 (2017)}\
  }\BibitemShut {NoStop}%
\bibitem [{\citenamefont {Hermanns}\ \emph {et~al.}(2018)\citenamefont
  {Hermanns}, \citenamefont {Kimchi},\ and\ \citenamefont
  {Knolle}}]{033117-053934}%
  \BibitemOpen
\bibfield  {journal} {  }\bibfield  {author} {\bibinfo {author} {\bibfnamefont
  {M.}~\bibnamefont {Hermanns}}, \bibinfo {author} {\bibfnamefont
  {I.}~\bibnamefont {Kimchi}},\ and\ \bibinfo {author} {\bibfnamefont
  {J.}~\bibnamefont {Knolle}},\ }\href@noop {} {\bibfield  {journal} {\bibinfo
  {journal} {Annual Review of Condensed Matter Physics}\ }\textbf {\bibinfo
  {volume} {9}},\ \bibinfo {pages} {17} (\bibinfo {year} {2018})}\BibitemShut
  {NoStop}%
\bibitem [{\citenamefont {Motome}\ and\ \citenamefont
  {Nasu}(2020)}]{doi:10.7566/JPSJ.89.012002}%
  \BibitemOpen
  \bibfield  {author} {\bibinfo {author} {\bibfnamefont {Y.}~\bibnamefont
  {Motome}}\ and\ \bibinfo {author} {\bibfnamefont {J.}~\bibnamefont {Nasu}},\
  }\href@noop {} {\bibfield  {journal} {\bibinfo  {journal} {Journal of the
  Physical Society of Japan}\ }\textbf {\bibinfo {volume} {89}},\ \bibinfo
  {pages} {012002} (\bibinfo {year} {2020})}\BibitemShut {NoStop}%
\bibitem [{\citenamefont {Kasahara}\ \emph {et~al.}(2018)\citenamefont
  {Kasahara}, \citenamefont {Ohnishi}, \citenamefont {Mizukami}, \citenamefont
  {Tanaka}, \citenamefont {Ma}, \citenamefont {Sugii}, \citenamefont {Kurita},
  \citenamefont {Tanaka}, \citenamefont {Nasu}, \citenamefont {Motome},
  \citenamefont {Shibauchi},\ and\ \citenamefont {Matsuda}}]{Kasahara2018}%
  \BibitemOpen
  \bibfield  {author} {\bibinfo {author} {\bibfnamefont {Y.}~\bibnamefont
  {Kasahara}}, \bibinfo {author} {\bibfnamefont {T.}~\bibnamefont {Ohnishi}},
  \bibinfo {author} {\bibfnamefont {Y.}~\bibnamefont {Mizukami}}, \bibinfo
  {author} {\bibfnamefont {O.}~\bibnamefont {Tanaka}}, \bibinfo {author}
  {\bibfnamefont {S.}~\bibnamefont {Ma}}, \bibinfo {author} {\bibfnamefont
  {K.}~\bibnamefont {Sugii}}, \bibinfo {author} {\bibfnamefont
  {N.}~\bibnamefont {Kurita}}, \bibinfo {author} {\bibfnamefont
  {H.}~\bibnamefont {Tanaka}}, \bibinfo {author} {\bibfnamefont
  {J.}~\bibnamefont {Nasu}}, \bibinfo {author} {\bibfnamefont {Y.}~\bibnamefont
  {Motome}}, \bibinfo {author} {\bibfnamefont {T.}~\bibnamefont {Shibauchi}},\
  and\ \bibinfo {author} {\bibfnamefont {Y.}~\bibnamefont {Matsuda}},\
  }\href@noop {} {\bibfield  {journal} {\bibinfo  {journal} {Nature}\ }\textbf
  {\bibinfo {volume} {559}},\ \bibinfo {pages} {227} (\bibinfo {year}
  {2018})}\BibitemShut {NoStop}%
\bibitem [{\citenamefont {Yokoi}\ \emph {et~al.}(2020)\citenamefont {Yokoi},
  \citenamefont {Ma}, \citenamefont {Kasahara}, \citenamefont {Kasahara},
  \citenamefont {Shibauchi}, \citenamefont {Kurita}, \citenamefont {Tanaka},
  \citenamefont {Nasu}, \citenamefont {Motome}, \citenamefont {Hickey},
  \citenamefont {Trebst},\ and\ \citenamefont
  {Matsuda}}]{yokoi2020halfinteger}%
  \BibitemOpen
  \bibfield  {author} {\bibinfo {author} {\bibfnamefont {T.}~\bibnamefont
  {Yokoi}}, \bibinfo {author} {\bibfnamefont {S.}~\bibnamefont {Ma}}, \bibinfo
  {author} {\bibfnamefont {Y.}~\bibnamefont {Kasahara}}, \bibinfo {author}
  {\bibfnamefont {S.}~\bibnamefont {Kasahara}}, \bibinfo {author}
  {\bibfnamefont {T.}~\bibnamefont {Shibauchi}}, \bibinfo {author}
  {\bibfnamefont {N.}~\bibnamefont {Kurita}}, \bibinfo {author} {\bibfnamefont
  {H.}~\bibnamefont {Tanaka}}, \bibinfo {author} {\bibfnamefont
  {J.}~\bibnamefont {Nasu}}, \bibinfo {author} {\bibfnamefont {Y.}~\bibnamefont
  {Motome}}, \bibinfo {author} {\bibfnamefont {C.}~\bibnamefont {Hickey}},
  \bibinfo {author} {\bibfnamefont {S.}~\bibnamefont {Trebst}},\ and\ \bibinfo
  {author} {\bibfnamefont {Y.}~\bibnamefont {Matsuda}},\ }\href@noop {} {}
  (\bibinfo {year} {2020}),\ \Eprint {https://arxiv.org/abs/2001.01899}
  {arXiv:2001.01899 [cond-mat.str-el]} \BibitemShut {NoStop}%
\bibitem [{\citenamefont {Tanaka}\ \emph {et~al.}(2020)\citenamefont {Tanaka},
  \citenamefont {Mizukami}, \citenamefont {Harasawa}, \citenamefont
  {Hashimoto}, \citenamefont {Kurita}, \citenamefont {Tanaka}, \citenamefont
  {Fujimoto}, \citenamefont {Matsuda}, \citenamefont {Moon},\ and\
  \citenamefont {Shibauchi}}]{tanaka2020thermodynamic}%
  \BibitemOpen
  \bibfield  {author} {\bibinfo {author} {\bibfnamefont {O.}~\bibnamefont
  {Tanaka}}, \bibinfo {author} {\bibfnamefont {Y.}~\bibnamefont {Mizukami}},
  \bibinfo {author} {\bibfnamefont {R.}~\bibnamefont {Harasawa}}, \bibinfo
  {author} {\bibfnamefont {K.}~\bibnamefont {Hashimoto}}, \bibinfo {author}
  {\bibfnamefont {N.}~\bibnamefont {Kurita}}, \bibinfo {author} {\bibfnamefont
  {H.}~\bibnamefont {Tanaka}}, \bibinfo {author} {\bibfnamefont
  {S.}~\bibnamefont {Fujimoto}}, \bibinfo {author} {\bibfnamefont
  {Y.}~\bibnamefont {Matsuda}}, \bibinfo {author} {\bibfnamefont {E.~G.}\
  \bibnamefont {Moon}},\ and\ \bibinfo {author} {\bibfnamefont
  {T.}~\bibnamefont {Shibauchi}},\ }\href@noop {} {} (\bibinfo {year} {2020}),\
  \Eprint {https://arxiv.org/abs/2007.06757} {arXiv:2007.06757
  [cond-mat.str-el]} \BibitemShut {NoStop}%
\bibitem [{Ban({\natexlab{a}})}]{Banerjee2016}%
  \BibitemOpen
  \href@noop {} {\bibfield  {journal} {\bibinfo  {journal} {A. Banerjee, C. A.
  Bridges, J. Q. Yan, A. A. Aczel, L. Li, M. B. Stone, G. E. Granroth, M. D.
  Lumsden, Y. Yiu, J. Knolle, S. Bhattacharjee, D. L. Kovrizhin, R. Moessner,
  D. A. Tennant, D. G. Mandrus, and S. E. Nagler, Nat. Mater. \textbf{15}, 733
  (2016)}\ } ({\natexlab{a}})}\BibitemShut {NoStop}%
\bibitem [{Ban({\natexlab{b}})}]{Banerjee2016a}%
  \BibitemOpen
  \href@noop {} {\bibfield  {journal} {\bibinfo  {journal} {A. Banerjee, J.
  Yan, J. Knolle, C. A. Bridges, M. B. Stone, M. D. Lumsden, D. G. Mandrus, D.
  A. Tennant, R. Moessner, and S. E. Nagler, Science \textbf{356}, 6342
  (2017)}\ } ({\natexlab{b}})}\BibitemShut {NoStop}%
\bibitem [{Ban({\natexlab{c}})}]{Banerjee2018}%
  \BibitemOpen
  \href@noop {} {\bibfield  {journal} {\bibinfo  {journal} {A. Banerjee, P.
  Lampen-Kelley, J. Knolle, C. Balz, A. A. Aczel, B. Winn, Y. Liu, D.
  Pajerowski, J. Yan, C. A. Bridges, A. T. Savici, B. C. Chakoumakos, M. D.
  Lumsden, D. A. Tennant, R. Moessner, D. G. Mandrus, and S. E. Nagler, Nat.
  Part. J. Quantum Mater. \textbf{3}, 8 (2018)}\ }
  ({\natexlab{c}})}\BibitemShut {NoStop}%
\bibitem [{\citenamefont {Knolle}\ \emph
  {et~al.}(2014{\natexlab{a}})\citenamefont {Knolle}, \citenamefont
  {Kovrizhin}, \citenamefont {Chalker},\ and\ \citenamefont
  {Moessner}}]{knolle2014dynamics}%
  \BibitemOpen
  \bibfield  {author} {\bibinfo {author} {\bibfnamefont {J.}~\bibnamefont
  {Knolle}}, \bibinfo {author} {\bibfnamefont {D.}~\bibnamefont {Kovrizhin}},
  \bibinfo {author} {\bibfnamefont {J.}~\bibnamefont {Chalker}},\ and\ \bibinfo
  {author} {\bibfnamefont {R.}~\bibnamefont {Moessner}},\ }\href@noop {}
  {\bibfield  {journal} {\bibinfo  {journal} {Physical Review Letters}\
  }\textbf {\bibinfo {volume} {112}},\ \bibinfo {pages} {207203} (\bibinfo
  {year} {2014}{\natexlab{a}})}\BibitemShut {NoStop}%
\bibitem [{\citenamefont {Smith}\ \emph {et~al.}(2015)\citenamefont {Smith},
  \citenamefont {Knolle}, \citenamefont {Kovrizhin}, \citenamefont {Chalker},\
  and\ \citenamefont {Moessner}}]{smith2015neutron}%
  \BibitemOpen
  \bibfield  {author} {\bibinfo {author} {\bibfnamefont {A.}~\bibnamefont
  {Smith}}, \bibinfo {author} {\bibfnamefont {J.}~\bibnamefont {Knolle}},
  \bibinfo {author} {\bibfnamefont {D.}~\bibnamefont {Kovrizhin}}, \bibinfo
  {author} {\bibfnamefont {J.}~\bibnamefont {Chalker}},\ and\ \bibinfo {author}
  {\bibfnamefont {R.}~\bibnamefont {Moessner}},\ }\href@noop {} {\bibfield
  {journal} {\bibinfo  {journal} {Physical Review B}\ }\textbf {\bibinfo
  {volume} {92}},\ \bibinfo {pages} {180408} (\bibinfo {year}
  {2015})}\BibitemShut {NoStop}%
\bibitem [{\citenamefont {Knolle}\ \emph {et~al.}(2018)\citenamefont {Knolle},
  \citenamefont {Bhattacharjee},\ and\ \citenamefont
  {Moessner}}]{knolle2018dynamics}%
  \BibitemOpen
  \bibfield  {author} {\bibinfo {author} {\bibfnamefont {J.}~\bibnamefont
  {Knolle}}, \bibinfo {author} {\bibfnamefont {S.}~\bibnamefont
  {Bhattacharjee}},\ and\ \bibinfo {author} {\bibfnamefont {R.}~\bibnamefont
  {Moessner}},\ }\href@noop {} {\bibfield  {journal} {\bibinfo  {journal}
  {Physical Review B}\ }\textbf {\bibinfo {volume} {97}},\ \bibinfo {pages}
  {134432} (\bibinfo {year} {2018})}\BibitemShut {NoStop}%
\bibitem [{Bae()}]{Baek2017}%
  \BibitemOpen
  \href@noop {} {\bibinfo  {journal} {S.-H. Baek, S.-H. Do, K. Y. Choi, Y.S.
  Kwon, A.U.B. Wolter, S. Nishimoto, J. van den Brink, and B. B{\"u}chner,
  Phys. Rev. Lett. \textbf{119}, 037201 (2017)}\ }\BibitemShut {NoStop}%
\bibitem [{Zhe()}]{Zheng2017}%
  \BibitemOpen
\bibfield  {journal} {  }\href@noop {} {\bibinfo  {journal} {J. Zheng, K. Ran,
  T. Li, J. Wang, P. Wang, B. Liu, Z.-X. Liu, B. Normand, J. Wen, and W. Yu,
  Phys. Rev. Lett. \textbf{119}, 227208 (2017)}\ }\BibitemShut {NoStop}%
\bibitem [{\citenamefont {Knolle}\ \emph
  {et~al.}(2014{\natexlab{b}})\citenamefont {Knolle}, \citenamefont {Chern},
  \citenamefont {Kovrizhin}, \citenamefont {Moessner},\ and\ \citenamefont
  {Perkins}}]{PhysRevLett.113.187201}%
  \BibitemOpen
\bibfield  {journal} {  }\bibfield  {author} {\bibinfo {author} {\bibfnamefont
  {J.}~\bibnamefont {Knolle}}, \bibinfo {author} {\bibfnamefont {G.-W.}\
  \bibnamefont {Chern}}, \bibinfo {author} {\bibfnamefont {D.~L.}\ \bibnamefont
  {Kovrizhin}}, \bibinfo {author} {\bibfnamefont {R.}~\bibnamefont
  {Moessner}},\ and\ \bibinfo {author} {\bibfnamefont {N.~B.}\ \bibnamefont
  {Perkins}},\ }\href@noop {} {\bibfield  {journal} {\bibinfo  {journal} {Phys.
  Rev. Lett.}\ }\textbf {\bibinfo {volume} {113}},\ \bibinfo {pages} {187201}
  (\bibinfo {year} {2014}{\natexlab{b}})}\BibitemShut {NoStop}%
\bibitem [{\citenamefont {Nasu}\ \emph {et~al.}(2016)\citenamefont {Nasu},
  \citenamefont {Knolle}, \citenamefont {Kovrizhin}, \citenamefont {Motome},\
  and\ \citenamefont {Moessner}}]{Nasu2016}%
  \BibitemOpen
  \bibfield  {author} {\bibinfo {author} {\bibfnamefont {J.}~\bibnamefont
  {Nasu}}, \bibinfo {author} {\bibfnamefont {J.}~\bibnamefont {Knolle}},
  \bibinfo {author} {\bibfnamefont {D.~L.}\ \bibnamefont {Kovrizhin}}, \bibinfo
  {author} {\bibfnamefont {Y.}~\bibnamefont {Motome}},\ and\ \bibinfo {author}
  {\bibfnamefont {R.}~\bibnamefont {Moessner}},\ }\href@noop {} {\bibfield
  {journal} {\bibinfo  {journal} {Nature Physics}\ }\textbf {\bibinfo {volume}
  {12}},\ \bibinfo {pages} {912} (\bibinfo {year} {2016})}\BibitemShut
  {NoStop}%
\bibitem [{\citenamefont {Perreault}\ \emph {et~al.}(2015)\citenamefont
  {Perreault}, \citenamefont {Knolle}, \citenamefont {Perkins},\ and\
  \citenamefont {Burnell}}]{perreault2015theory}%
  \BibitemOpen
  \bibfield  {author} {\bibinfo {author} {\bibfnamefont {B.}~\bibnamefont
  {Perreault}}, \bibinfo {author} {\bibfnamefont {J.}~\bibnamefont {Knolle}},
  \bibinfo {author} {\bibfnamefont {N.~B.}\ \bibnamefont {Perkins}},\ and\
  \bibinfo {author} {\bibfnamefont {F.}~\bibnamefont {Burnell}},\ }\href@noop
  {} {\bibfield  {journal} {\bibinfo  {journal} {Physical Review B}\ }\textbf
  {\bibinfo {volume} {92}},\ \bibinfo {pages} {094439} (\bibinfo {year}
  {2015})}\BibitemShut {NoStop}%
\bibitem [{\citenamefont {Vinkler-Aviv}\ and\ \citenamefont
  {Rosch}(2018)}]{PhysRevX.8.031032}%
  \BibitemOpen
  \bibfield  {author} {\bibinfo {author} {\bibfnamefont {Y.}~\bibnamefont
  {Vinkler-Aviv}}\ and\ \bibinfo {author} {\bibfnamefont {A.}~\bibnamefont
  {Rosch}},\ }\href@noop {} {\bibfield  {journal} {\bibinfo  {journal} {Phys.
  Rev. X}\ }\textbf {\bibinfo {volume} {8}},\ \bibinfo {pages} {031032}
  (\bibinfo {year} {2018})}\BibitemShut {NoStop}%
\bibitem [{\citenamefont {Ye}\ \emph {et~al.}(2018)\citenamefont {Ye},
  \citenamefont {Hal\'asz}, \citenamefont {Savary},\ and\ \citenamefont
  {Balents}}]{PhysRevLett.121.147201}%
  \BibitemOpen
  \bibfield  {author} {\bibinfo {author} {\bibfnamefont {M.}~\bibnamefont
  {Ye}}, \bibinfo {author} {\bibfnamefont {G.~B.}\ \bibnamefont {Hal\'asz}},
  \bibinfo {author} {\bibfnamefont {L.}~\bibnamefont {Savary}},\ and\ \bibinfo
  {author} {\bibfnamefont {L.}~\bibnamefont {Balents}},\ }\href
  {https://doi.org/10.1103/PhysRevLett.121.147201} {\bibfield  {journal}
  {\bibinfo  {journal} {Phys. Rev. Lett.}\ }\textbf {\bibinfo {volume} {121}},\
  \bibinfo {pages} {147201} (\bibinfo {year} {2018})}\BibitemShut {NoStop}%
\bibitem [{\citenamefont {Metavitsiadis}\ and\ \citenamefont
  {Brenig}(2020{\natexlab{b}})}]{PhysRevB.101.035103}%
  \BibitemOpen
  \bibfield  {author} {\bibinfo {author} {\bibfnamefont {A.}~\bibnamefont
  {Metavitsiadis}}\ and\ \bibinfo {author} {\bibfnamefont {W.}~\bibnamefont
  {Brenig}},\ }\href@noop {} {\bibfield  {journal} {\bibinfo  {journal} {Phys.
  Rev. B}\ }\textbf {\bibinfo {volume} {101}},\ \bibinfo {pages} {035103}
  (\bibinfo {year} {2020}{\natexlab{b}})}\BibitemShut {NoStop}%
\bibitem [{\citenamefont {Ye}\ \emph {et~al.}(2020)\citenamefont {Ye},
  \citenamefont {Fernandes},\ and\ \citenamefont
  {Perkins}}]{PhysRevResearch.2.033180}%
  \BibitemOpen
  \bibfield  {author} {\bibinfo {author} {\bibfnamefont {M.}~\bibnamefont
  {Ye}}, \bibinfo {author} {\bibfnamefont {R.~M.}\ \bibnamefont {Fernandes}},\
  and\ \bibinfo {author} {\bibfnamefont {N.~B.}\ \bibnamefont {Perkins}},\
  }\href@noop {} {\bibfield  {journal} {\bibinfo  {journal} {Phys. Rev.
  Research}\ }\textbf {\bibinfo {volume} {2}},\ \bibinfo {pages} {033180}
  (\bibinfo {year} {2020})}\BibitemShut {NoStop}%
\bibitem [{Li2()}]{Li2020}%
  \BibitemOpen
  \href@noop {} {\bibinfo  {journal} {H. Li, T. T. Zhang, A. Said, G. Fabbris,
  D. G. Mazzone, J. Q. Yan, D. Mandrus, G. B. Halasz, S. Okamoto, S. Murakami,
  M. P. M. Dean, H. N. Lee, and H. Miao, {\it ArXiv:2011.07036 [Cond-Mat]}
  (2020)}\ }\BibitemShut {NoStop}%
\bibitem [{\citenamefont {Baskaran}\ \emph {et~al.}(2007)\citenamefont
  {Baskaran}, \citenamefont {Mandal},\ and\ \citenamefont
  {Shankar}}]{PhysRevLett.98.247201}%
  \BibitemOpen
\bibfield  {journal} {  }\bibfield  {author} {\bibinfo {author} {\bibfnamefont
  {G.}~\bibnamefont {Baskaran}}, \bibinfo {author} {\bibfnamefont
  {S.}~\bibnamefont {Mandal}},\ and\ \bibinfo {author} {\bibfnamefont
  {R.}~\bibnamefont {Shankar}},\ }\href
  {https://doi.org/10.1103/PhysRevLett.98.247201} {\bibfield  {journal}
  {\bibinfo  {journal} {Phys. Rev. Lett.}\ }\textbf {\bibinfo {volume} {98}},\
  \bibinfo {pages} {247201} (\bibinfo {year} {2007})}\BibitemShut {NoStop}%
\bibitem [{\citenamefont {Metavitsiadis}\ \emph {et~al.}(2017)\citenamefont
  {Metavitsiadis}, \citenamefont {Pidatella},\ and\ \citenamefont
  {Brenig}}]{PhysRevB.96.205121}%
  \BibitemOpen
  \bibfield  {author} {\bibinfo {author} {\bibfnamefont {A.}~\bibnamefont
  {Metavitsiadis}}, \bibinfo {author} {\bibfnamefont {A.}~\bibnamefont
  {Pidatella}},\ and\ \bibinfo {author} {\bibfnamefont {W.}~\bibnamefont
  {Brenig}},\ }\href@noop {} {\bibfield  {journal} {\bibinfo  {journal} {Phys.
  Rev. B}\ }\textbf {\bibinfo {volume} {96}},\ \bibinfo {pages} {205121}
  (\bibinfo {year} {2017})}\BibitemShut {NoStop}%
\bibitem [{\citenamefont {Pidatella}\ \emph {et~al.}(2019)\citenamefont
  {Pidatella}, \citenamefont {Metavitsiadis},\ and\ \citenamefont
  {Brenig}}]{PhysRevB.99.075141}%
  \BibitemOpen
  \bibfield  {author} {\bibinfo {author} {\bibfnamefont {A.}~\bibnamefont
  {Pidatella}}, \bibinfo {author} {\bibfnamefont {A.}~\bibnamefont
  {Metavitsiadis}},\ and\ \bibinfo {author} {\bibfnamefont {W.}~\bibnamefont
  {Brenig}},\ }\href@noop {} {\bibfield  {journal} {\bibinfo  {journal} {Phys.
  Rev. B}\ }\textbf {\bibinfo {volume} {99}},\ \bibinfo {pages} {075141}
  (\bibinfo {year} {2019})}\BibitemShut {NoStop}%
\bibitem [{sup()}]{supplement}%
  \BibitemOpen
  \href@noop {} {}\bibinfo {note} {See supplemental material at
  \href{www.aps.org}{www.aps.org}.}\BibitemShut {Stop}%
\bibitem [{\citenamefont {Rau}\ \emph {et~al.}(2014)\citenamefont {Rau},
  \citenamefont {Lee},\ and\ \citenamefont {Kee}}]{Rau2014}%
  \BibitemOpen
  \bibfield  {author} {\bibinfo {author} {\bibfnamefont {J.~G.}\ \bibnamefont
  {Rau}}, \bibinfo {author} {\bibfnamefont {E.~K.-H.}\ \bibnamefont {Lee}},\
  and\ \bibinfo {author} {\bibfnamefont {H.-Y.}\ \bibnamefont {Kee}},\ }\href
  {https://doi.org/10.1103/PhysRevLett.112.077204} {\bibfield  {journal}
  {\bibinfo  {journal} {Phys. Rev. Lett.}\ }\textbf {\bibinfo {volume} {112}},\
  \bibinfo {pages} {077204} (\bibinfo {year} {2014})}\BibitemShut {NoStop}%
\bibitem [{\citenamefont {Natori}\ \emph {et~al.}(2019)\citenamefont {Natori},
  \citenamefont {Moessner},\ and\ \citenamefont {Knolle}}]{Natori2019A}%
  \BibitemOpen
  \bibfield  {author} {\bibinfo {author} {\bibfnamefont {W.~M.~H.}\
  \bibnamefont {Natori}}, \bibinfo {author} {\bibfnamefont {R.}~\bibnamefont
  {Moessner}},\ and\ \bibinfo {author} {\bibfnamefont {J.}~\bibnamefont
  {Knolle}},\ }\href {https://doi.org/10.1103/PhysRevB.100.144403} {\bibfield
  {journal} {\bibinfo  {journal} {Phys. Rev. B}\ }\textbf {\bibinfo {volume}
  {100}},\ \bibinfo {pages} {144403} (\bibinfo {year} {2019})}\BibitemShut
  {NoStop}%
\bibitem [{\citenamefont {Kugel}\ and\ \citenamefont
  {Khomskii}(1982)}]{Kugel1982}%
  \BibitemOpen
  \bibfield  {author} {\bibinfo {author} {\bibfnamefont {K.~I.}\ \bibnamefont
  {Kugel}}\ and\ \bibinfo {author} {\bibfnamefont {D.~I.}\ \bibnamefont
  {Khomskii}},\ }\href@noop {} {\bibfield  {journal} {\bibinfo  {journal} {Sov.
  Phys. Usp.}\ }\textbf {\bibinfo {volume} {25}},\ \bibinfo {pages} {231}
  (\bibinfo {year} {1982})}\BibitemShut {NoStop}%
\bibitem [{\citenamefont {Fano}(1961)}]{PhysRev.124.1866}%
  \BibitemOpen
  \bibfield  {author} {\bibinfo {author} {\bibfnamefont {U.}~\bibnamefont
  {Fano}},\ }\href@noop {} {\bibfield  {journal} {\bibinfo  {journal} {Phys.
  Rev.}\ }\textbf {\bibinfo {volume} {124}},\ \bibinfo {pages} {1866} (\bibinfo
  {year} {1961})}\BibitemShut {NoStop}%
\bibitem [{\citenamefont {Fleury}\ and\ \citenamefont
  {Loudon}(1968)}]{PhysRev.166.514}%
  \BibitemOpen
  \bibfield  {author} {\bibinfo {author} {\bibfnamefont {P.~A.}\ \bibnamefont
  {Fleury}}\ and\ \bibinfo {author} {\bibfnamefont {R.}~\bibnamefont
  {Loudon}},\ }\href@noop {} {\bibfield  {journal} {\bibinfo  {journal} {Phys.
  Rev.}\ }\textbf {\bibinfo {volume} {166}},\ \bibinfo {pages} {514} (\bibinfo
  {year} {1968})}\BibitemShut {NoStop}%
\bibitem [{\citenamefont {Kaib}\ \emph {et~al.}(2020)\citenamefont {Kaib},
  \citenamefont {Biswas}, \citenamefont {Riedl}, \citenamefont {Winter},\ and\
  \citenamefont {Valenti}}]{kaib2020magnetoelastic}%
  \BibitemOpen
  \bibfield  {author} {\bibinfo {author} {\bibfnamefont {D.}~\bibnamefont
  {Kaib}}, \bibinfo {author} {\bibfnamefont {S.}~\bibnamefont {Biswas}},
  \bibinfo {author} {\bibfnamefont {K.}~\bibnamefont {Riedl}}, \bibinfo
  {author} {\bibfnamefont {S.}~\bibnamefont {Winter}},\ and\ \bibinfo {author}
  {\bibfnamefont {R.}~\bibnamefont {Valenti}},\ }\href@noop {} {\bibfield
  {journal} {\bibinfo  {journal} {arXiv preprint arXiv:2008.08616}\ } (\bibinfo
  {year} {2020})}\BibitemShut {NoStop}%
\end{thebibliography}
\end{document}